\documentclass[ALICE,manyauthors]{cernphprep}

\usepackage[comma,square,numbers,sort&compress]{natbib}
\usepackage{hyperref}
\usepackage{lineno}
\def\snn{\mbox{$\sqrt{s_{_{\rm NN}}}$}}
\def \pb{{Pb--Pb}}

\newcommand{ \be }{\begin{eqnarray}}
\newcommand{ \ee }{\end{eqnarray}}

\newcommand{ \bk }{{\bf k}}
\newcommand{ \bp }{{\bf p}}

\newcommand{ \bq }{{\bf q}}

\newcommand{ \rT }{{\rm T}}
\newcommand{ \eps }{\varepsilon}
\newcommand{ \mean }[1]{\left\langle #1 \right\rangle}

\newcommand{\pt}{$p_{\rm{T}}$~}
\newcommand{\kt}{k_{\rm{T}}}
\newcommand{\gevc}{{\rm GeV/{\it c}}}
\newcommand{\rside}{R_{\rm side}}
\newcommand{\rout}{R_{\rm out}}
\newcommand{\rlong}{R_{\rm long}}
\newcommand{\ros}{R_{\rm os}}
\newcommand{\rol}{R_{\rm ol}}
\newcommand{\rsl}{R_{\rm sl}}
\newcommand{\qside}{q_{\rm side}}
\newcommand{\qout}{q_{\rm out}}
\newcommand{\qlong}{q_{\rm long}}
\newcommand{\rsidez}{R_{{\rm side},0}}
\newcommand{\routz}{R_{{\rm out},0}}
\newcommand{\rlongz}{R_{{\rm long},0}}
\newcommand{\rosz}{R_{{\rm os},0}}
\newcommand{\rsidetwo}{R_{{\rm side},2}}

 \newcommand{\dphi}{\Delta\varphi}
\newcommand{\PsiT}{\Psi_{\mathrm{EP},2}}
\makeatletter
\def\NAT@def@citea{\def\@citea{\NAT@separator}}
\makeatother

\begin{document}%

%%%%%%%%%%%%%%%  Title page %%%%%%%%%%%%%%%%%%%%%%%%
\begin{titlepage}
\PHyear{2017}
\PHnumber{013}      % required, will be obtained from PH
\PHdate{January}  % required, will be obtained from PH
%

%%% Put your own title + short title here:
\title{Azimuthally differential pion femtoscopy\\ in \pb~collisions
  at \snn~= 2.76~TeV }

\ShortTitle{Azimuthally differential pion femtoscopy in
  \pb~collisions}   % appears on right page headers

%%% Do not change the next lines
\Collaboration{ALICE Collaboration\thanks{See
    Appendix~\ref{app:collab} for the list of collaboration members}}
\ShortAuthor{ALICE Collaboration}
% appears on left page headers, do not change

\begin{abstract}
We present the first azimuthally differential measurements of the pion
source size relative to the second harmonic event plane in
\pb~collisions at a center-of-mass energy per nucleon-nucleon pair of
\snn~= 2.76~TeV. The measurements have been performed in the
centrality range 0--50\% and for pion pair transverse momenta $0.2 <
\kt < 0.7$~\gevc. We find that the $\rside$ and $\rout$ radii, which
characterize the pion source size in the directions perpendicular and
parallel to the pion transverse momentum, oscillate out of phase,
similar to what was observed at the Relativistic Heavy Ion Collider
(RHIC). The final-state source eccentricity, estimated via $\rside$
oscillations, is found to be significantly smaller than the
initial-state source eccentricity, but remains positive -- indicating
that even after a stronger expansion in the in-plane direction, the
pion source at the freeze-out is still elongated in the out-of-plane
direction. The 3+1D hydrodynamic calculations are in qualitative agreement with
observed centrality and transverse momentum $\rside$ oscillations, but
systematically underestimate the oscillation magnitude.
\end{abstract}
\end{titlepage}
\setcounter{page}{2}

\maketitle

%\linenumbers

It was first shown in 1960 that the distribution of
pions emitted in ${\rm p}\bar{{\rm p}}$ collisions at small relative angles is
affected by quantum statistical effects and is sensitive to the size
of the emitting source~\cite{Goldhaber:1960sf}. Since then, the correlation technique with two
identical particles at small relative momentum, often called
intensity, or Hanbury Brown--Twiss (HBT),
interferometry~\cite{Kopylov:1972qw,Kopylov:1973qq,Kopylov:1974th,Pratt:1986cc,Bertsch:1988db},
has been used to study the space-time structure of the pion-emitting
source from hadron-hadron and electron-positron to heavy-ion collisions (for a
review, see~\cite{Lisa:2005dd}).  The so-called HBT radii, obtained in
these analyses, characterize the spatial and temporal extent of the
source emitting pions of a given momentum, the extensions of the
so-called homogeneity regions. Due to the position-momentum correlations
in particle emission, the HBT radii become sensitive to the collective
velocity fields, and as such provide information on the dynamics of
the system evolution~\cite{Lisa:2005dd}. Recent measurements of the centrality
dependence of the HBT radii in \pb~collisions at LHC energies~\cite{Aamodt:2010cz} further
confirm the scaling of the effective source volume with the particle
rapidity density as well as stronger radial flow at higher energies.

Pion interferometry of anisotropic sources (azimuthally differential
femtoscopy) was suggested in~\cite{Voloshin:1995mc,Voloshin:1996ch}, and the
corresponding measurements~\cite{Miskowiec:1995df} appeared shortly
after strong directed and in-plane elliptic flow were measured in
Au--Au collisions at the Alternating Gradient Synchrotron
(AGS)~\cite{Barrette:1994xr,Barrette:1996rs}.  Anisotropic flow, the
response of the system to the initial geometry, is usually
characterized by the Fourier decomposition of the particle azimuthal
distribution and quantified by the harmonic strength and orientation
of the corresponding flow plane.  Azimuthally differential femtoscopic
measurements can be performed relative to different harmonic flow planes, providing important complementary information on the
particle source. For example, the measurements of HBT radii with respect to the first
harmonic (directed) flow at the AGS~\cite{Lisa:2000xj} revealed that the source was tilted relative to the beam direction~\cite{Lisa:2000ip}. Azimuthal dependence of the HBT radii relative to the
higher harmonic ($n>2$) flow planes can originate only from the
anisotropies in collective flow
gradients~\cite{Voloshin:2011mg,Plumberg:2013nga} and the
observation~\cite{Adare:2014vax} of such a modulation unambiguously
signals a collective expansion and anisotropy in the flow fields. In particular, measurements of HBT radii with respect to the second harmonic (elliptic) flow provide information on the evolution of the system shape,
which is expected to become more spherical at freeze-out compared to
the initial state due to stronger in-plane expansion.  In the recent
RHIC beam energy scan, it was found that the eccentricity at
freeze-out decreases continuously with increasing beam
energy~\cite{Adamczyk:2014mxp}, a trend consistent with predictions by
hydrodynamic and hadronic transport models~\cite{Shen:2012us,Lisa:2011na}. Earlier
measurements~\cite{Adams:2003ra,Adams:2004yc} showed that even at the highest RHIC energies the source at
freeze-out remains out-of-plane extended, albeit with eccentricities
significantly lower than the initial ones. Hydrodynamical
calculations~\cite{Shen:2012us} predicted that at the Large Hadron
Collider (LHC) energies, about an order of magnitude higher than the
top RHIC energy, the pion source should eventually become isotropic,
or even in-plane extended.

In this Letter, we present the first azimuthally differential
femtoscopic measurements relative to the second harmonic flow
plane in \pb~collisions at \snn~= 2.76~TeV from the ALICE experiment
at the CERN-LHC and compare the results to previous measurements at
RHIC energies and to model calculations.

%==========================================================
%%\section{Experiment and data analysis}

The data were recorded in 2011 during the second~\pb\ running period
of the LHC. Approximately 2 million minimum bias
events, 29.2 million central trigger events, and 34.1 million
semi-central trigger events were used in this analysis.  A detailed description of the ALICE
detector can be found in~\cite{ALICE_exp_1,Abelev:2014ffa}.
The Time Projection Chamber (TPC) has full azimuthal coverage and allows
charged-particle track reconstruction in the pseudorapidity
range $|\eta|<~0.8$, as well as particle identification
via the specific ionization energy loss $\mathrm{d}E/\mathrm{d}x$ associated with each track.  In addition to the TPC, the
Time-Of-Flight (TOF) detector was used for identification of particles
with transverse momentum~\pt~$>$~0.5~\gevc.

The minimum bias, semi-central, and central triggers used in this analysis all require a signal in both V0
detectors~\cite{Abbas:2013taa}.  The V0 is a small angle detector
of scintillator arrays covering pseudorapidity ranges $2.8<\eta<5.1$
and $-3.7<\eta<-1.7$ for a collision vertex occurring at the center of
the ALICE detector.  The V0 detector was also used for the
centrality determination~\cite{Aamodt:2010cz}. The results of this analysis
are reported for collision centrality classes expressed as ranges of
the fraction of the inelastic \pb~cross-section: 0--5\%, 5--10\%,
10--20\%, 20--30\%, 30--40\%, and 40--50\%. The position of the
primary event vertex along the beam direction $V_{z}$ was determined
for each event.  Events with $|V_{z}|<8$~cm were used in this analysis
to ensure a uniform pseudorapidity acceptance.

The TPC has 18 sectors covering full azimuth with 159 pad rows
radially placed in each sector. Tracks with at least 80 space points
in the TPC have been used in this analysis.  Tracks compatible with a decay in flight (kink topology) were rejected.
The track quality was determined
by the $\chi^{2}$ of the Kalman filter fit to the reconstructed TPC
clusters.  The $\chi^2$ per degrees of freedom was
required to be less than 4.  For primary track selection, only
trajectories passing within 3.2~cm from the primary vertex in the
longitudinal direction and 2.4~cm in the transverse direction were
used.  Based on the specific ionization energy loss in the TPC gas
compared with the corresponding Bethe-Bloch curve, and the time of
flight in TOF, a probability for each track to be a pion, kaon,
proton, or electron was determined. Particles for which the pion
probability was the largest were used in this analysis.  Pions were
selected in the pseudorapidity range $|\eta|<0.8$ and 0.15 $<$ \pt$<$
1.5~\gevc.

%=======================================================================
%\section{Correlation Function}

The correlation function $C({\bq})$ was calculated as
\begin{eqnarray}
  C({\bq})=\frac{ A({\bq})} { B({\bq})},
\end{eqnarray}
where $\bq=\bp_1-\bp_2$ is the relative momentum of two pions,
$A(\bq)$ is the same-event distribution of particle pairs, and
$B(\bq)$ is the background distribution of uncorrelated particle
pairs. Both the $A({\bq})$ and $B({\bq})$ distributions were measured
differentially with respect to the second harmonic event-plane angle
$\PsiT$. The second harmonic event-plane angle $\PsiT$ was determined using TPC tracks. To avoid
self-correlation, each event was split into two subevents
($-0.8<\eta<0$ and $0<\eta<0.8$). Pairs were chosen from one subevent
and the second harmonic event-plane angle $\PsiT$ was determined using
the other subevent particles, and vice-versa, with the event plane
resolution determined from the correlations between the event planes
determined in different subevents~\cite{Poskanzer:1998yz}.
The background distribution is built by using the mixed-event
technique~\cite{Kopylov:1974th} in which pairs are made out of
particles from two different events with similar centrality (less than
2\% difference), event-plane angle (less than 10$^\circ$ difference),
and event vertex position along the beam direction (less than 4~cm
difference).  Requiring a minimum value in the
two-track separation parameters $\Delta \varphi^*$ and $\Delta\eta$
controls two-track reconstruction effects such as track splitting or
track merging.  The quantity $\varphi^{*}$ is defined in this analysis
as the azimuthal angle of the track in the laboratory frame at the
radial position of 1.6~m inside the TPC. Splitting is the effect when
one track is reconstructed as two tracks, and merging is the effect of
two tracks being reconstructed as one. Also, to reduce the splitting
effect, pairs that share more than 5\% of the TPC clusters were
removed from the analysis. It is observed that at large relative
momentum the correlation function is a constant, and the background
pair distribution is normalized such that this constant is unity.  The
analysis was performed for different collision centralities in several
ranges of $k_\rT$, the magnitude of the pion-pair transverse momentum
$\bk_\rT=(\bp_{\rT,1}+\bp_{\rT,2})/2$, and in bins of $\dphi=
\mathrm{\varphi_{pair}}-\PsiT$, defined in the range (0,~$\pi$) where
$\mathrm{\varphi_{pair}}$ is the pair azimuthal angle.  The
Bertsch-Pratt~\cite{Pratt:1986cc,Bertsch:1988db} out--side--long
coordinate system was used with the {\it long} direction pointing
along the beam axis, {\it out} along the transverse pair momentum, and
{\it side} being perpendicular to the other two.  The
three-dimensional correlation function was analyzed in the
Longitudinally Co-Moving System (LCMS), in which the total
longitudinal momentum of the pair is zero,
${p_{1,\mathrm{L}}}=-{p_{2,\mathrm{L}}}$.

To isolate the Bose-Einstein contribution in the correlation function, effects due to final-state Coulomb
repulsion must be taken into account.  For that, the Bowler-Sinyukov fitting
procedure~\cite{Bowler:1991pi,Sinyukov:1998fc} was used in which the
Coulomb weight is only applied to the fraction of pairs ($\lambda$)
that participate in the Bose-Einstein correlation. In this approach,
the correlation function is fitted to
\begin{eqnarray}
 C({\bq},\dphi)=N[(1-\lambda)+\lambda K({\bq})(1+G({\bq},\dphi))],
\end{eqnarray}
where $N$ is the normalization factor. The function $G({\bq},\dphi)$
describes the Bose-Einstein correlations and $K({\bq})$ is the Coulomb
part of the two-pion wave function integrated over a source function
corresponding to $G({\bq})$.  In this analysis the Gaussian form of
$G({\bq},\dphi)$ was used~\cite{guassain}:
\begin{eqnarray}
 G(\bq,\dphi)=\exp
\left[
-\qout^{2} \rout^{2}(\dphi)-\qside^{2} \rside^{2}(\dphi) \right.
\nonumber
\\
\left.
-\qlong^{2} \rlong^{2}(\dphi)-2\qout \qside \ros^{2}(\dphi) \right.
\nonumber
\\
\left.
-2\qside \qlong \rsl^{2}(\dphi)-2\qout \qlong \rol^{2}(\dphi)
\right],
\end{eqnarray}
where the parameters $\rout$, $\rside$, and $\rlong$ are traditionally
called HBT radii in the {\it out}, {\it side}, and {\it long}
directions.  The cross-terms $\ros^{2}$, $\rsl^{2}$, and $\rol^{2}$
describe the correlation in the {\it out}-{\it side}, {\it side}-{\it
  long}, and {\it out}-{\it long}~directions, respectively.

The systematic errors on the extracted radii vary within 3--9\%
depending on $\kt$ and centrality.  They include uncertainties related
to the tracking efficiency and track quality, momentum resolution
~\cite{Adam:2015vna}, different pair cuts ($\Delta \varphi^*$ and
$\Delta\eta$), and correlation function fit ranges.  Positive and
negative pion pairs as well as
data obtained with two opposite magnetic field polarities of the ALICE
L3 magnet
%pion pairs from different magnetic field orientations
have been analyzed separately and a small
difference in the results (less than 3\%) has been also accounted for
in the systematic error. The total systematic errors were obtained
from adding the above systematic errors in quadrature.

Other than being differential in the event plane, this analysis is
similar in most aspects to the analysis reported
in~\cite{Adam:2015vna}, and further details can be found there.
The results reported below were obtained with the second harmonic
event plane~\cite{Poskanzer:1998yz} determined with the TPC tracks.
It was checked that they are consistent with the results obtained with
the event-plane angle determined with the V0 detector.

%=================================================================
%\section{Results}

\begin{figure}[!ht]
 \centering
\hspace*{-5mm}
\includegraphics[width=11cm]{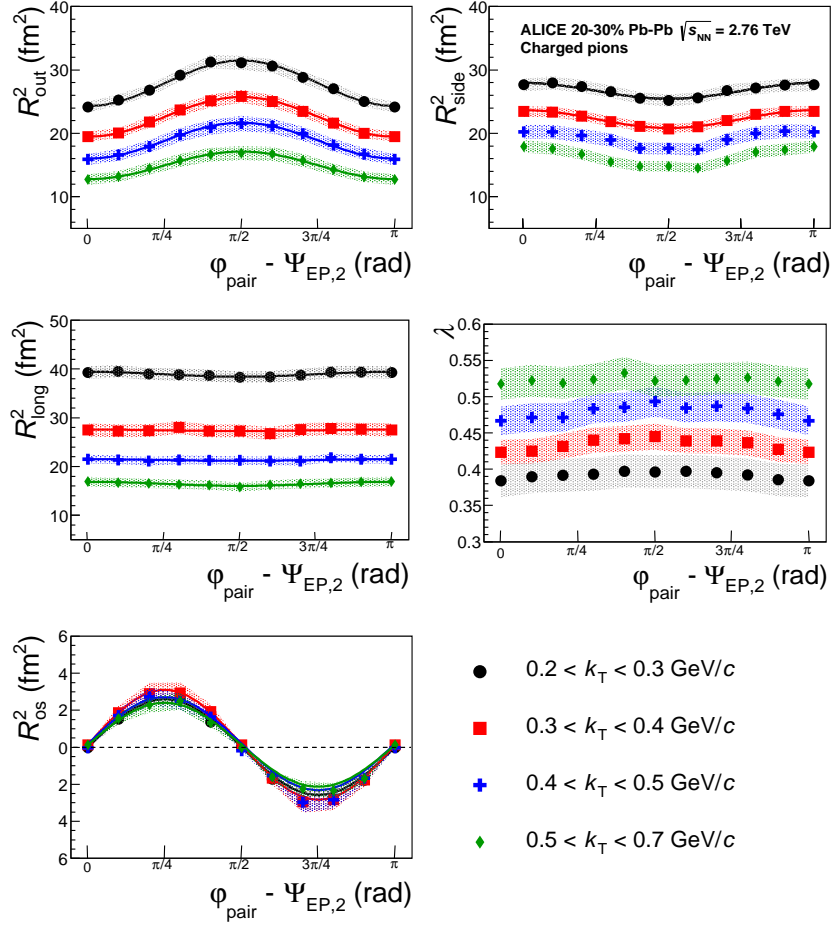}
\hspace*{-5mm}
\caption{The azimuthal dependence of $\rout^{2} $, $\rside^{2}$,
  $\rlong^{2}$, $\ros^{2}$, and
  $\lambda$ as a function of
  $\dphi=\mathrm{\varphi_{pair}}-\PsiT$ for the centrality 20--30\% and $k_\rT$
  ranges 0.2--0.3, 0.3--0.4, 0.4--0.5, and 0.5--0.7 GeV/$c$. Bands
  indicate the systematic errors. The results are not corrected for the
  event plane resolution of about 85--95\%.}
\label{fig:ktdependence}
\end{figure}

Figure~\ref{fig:ktdependence} presents the dependence of $\rout^{2}$,
$\rside^{2}$, $\rlong^{2}$, $\ros^{2}$, and $\lambda$ on the pion
emission angle relative to the second harmonic event plane. The
results are shown for the centrality classes 20--30\% in four ranges of
$\kt$: 0.2--0.3, 0.3--0.4, 0.4--0.5, and 0.5--0.7~\gevc.  $\rout^2$
and $\rside^2$ exhibit clear out-of-phase oscillations. No
oscillations for $\rlong^2$ and $\lambda$ are observed within the
uncertainties of the measurement. The parameter $\ros^2$ shows very
similar oscillations for all $\kt$ bins.  $\rol^2$ and $\rsl^2$ (not
shown) are found to be consistent with zero, as expected due to
symmetry, and are not further investigated in this analysis. A
possible correlation between $\lambda$ and the extracted radii was
checked by fixing $\lambda$. No change in the radii has been
observed. The curves represent the fits to the data using the
functions~\cite{Voloshin:1995mc,Voloshin:1996ch}:
 \begin{eqnarray} \label{eq:radii_osc}
 R^{2}_{\mu}(\dphi)=R^{2}_{\mu,0}+2R^{2}_{\mu,2}\cos(2\dphi)~(\mu={\rm out,side,long,sl,ol}),
\nonumber
\\
\nonumber
\ros^{2}(\dphi)=R^{2}_{{\rm os},0}+2 R^{2}_{{\rm os},2}\sin(2\dphi).
\end{eqnarray}
Fitting the radii's azimuthal dependence with the functional form of
Eq.~\ref{eq:radii_osc} allows us to extract the average radii and the
amplitudes of oscillations. The latter have to be corrected for the
finite event plane resolution. There exist several methods for such a
correction~\cite{Lisa:2005dd}, which produce very similar
results~\cite{Adamczyk:2014mxp} well within errors of this
analysis. The results shown below have been obtained with the simplest
method first used by the E895 Collaboration~\cite{Lisa:2000xj}, in
which the amplitude of oscillation is divided by the event plane
resolution factor.  The correction is about 5--15\%, depending
on centrality.
\begin{figure}[!ht]
\centering
\hspace*{-2mm}
\includegraphics[width=13cm]
{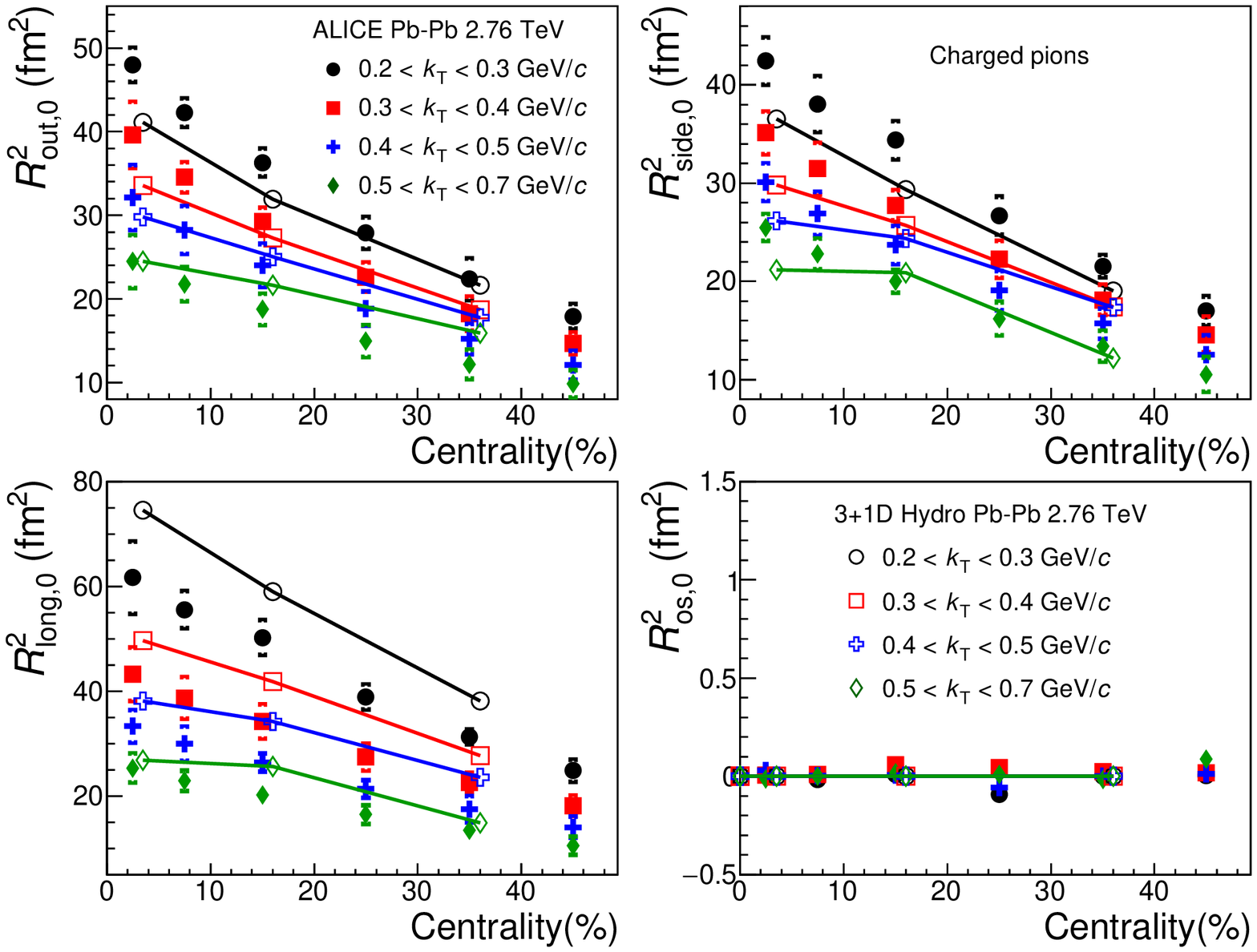}
  \caption{ The average radii $\routz^{2} $, $\rsidez^{2}$,
    $\rlongz^{2}$, and $\rosz^{2}$ as a function of centrality for
    different $\kt$ ranges compared to hydrodynamical
    calculations~\cite{Bozek:2014hwa}. Square brackets indicate the
    systematic errors.  }
\label{fig:avg_radii}
\end{figure}
Figure~\ref{fig:avg_radii} shows the average radii for different $\kt$
values as a function of centrality.  The average radii obtained in
this analysis are consistent with the results reported
in~\cite{Adam:2015vna}.  As expected, the radii are larger in more
central collisions and at smaller $\kt$ values, the latter reflecting
the effect of radial flow~\cite{Lisa:2005dd,Retiere:2003kf}.  The
cross-term $\rosz^2$ is consistent with zero, as expected due to the
symmetry of the system. Figure~\ref{fig:avg_radii} also shows the
average radii calculated for charged pions in the pseudorapidity range
$|\eta|<2$ from 3+1D hydrodynamic
calculations~\cite{Bozek:2014hwa}, assuming freeze-out temperature
$T_{f}$~=~150 MeV and a constant shear viscosity to entropy density
ratio $\eta/s$~= 0.08. The 3+1D hydrodynamic calculations,
while correctly describing the qualitative features of the average
radii dependence on centrality and $\kt$, fail to describe our results
quantitatively.
\begin{figure}[!ht]
\hspace*{14mm}
\includegraphics[width=13cm,keepaspectratio]{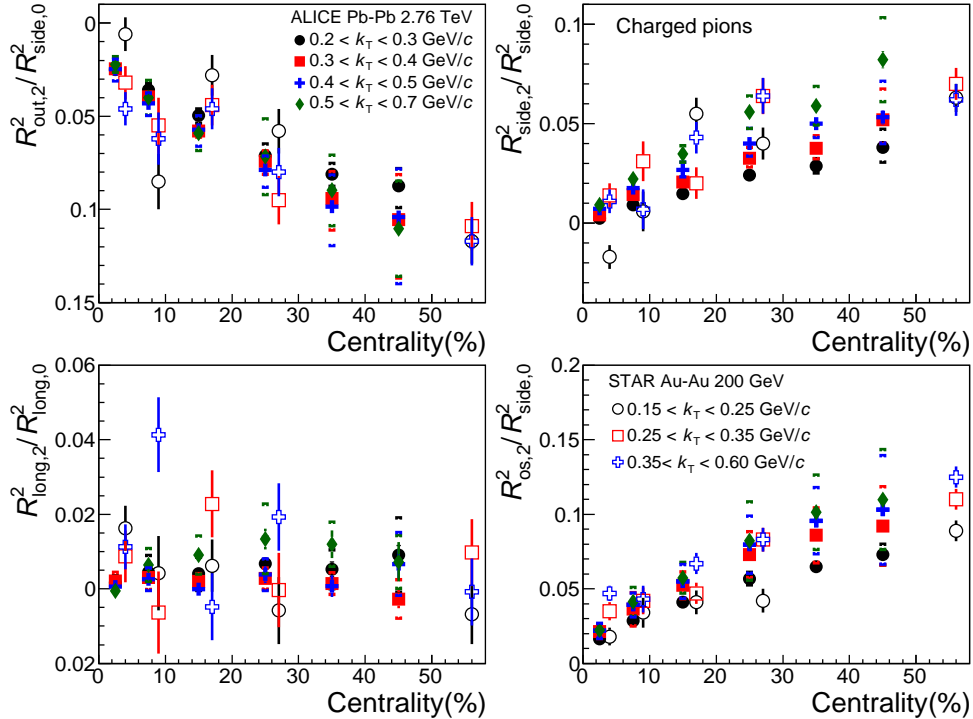}
  \caption{ Amplitudes of the relative radius oscillations~$R^{2}_{\rm
      out,2}/R^{2}_{\rm side,0}$,~$R^{2}_{\rm side,2}/R^{2}_{\rm
      side,0}$,~$R^{2}_{\rm long,2}/R^{2}_{\rm long,0}$,
    and~$R^{2}_{\rm os,2}/R^{2}_{\rm side,0}$ versus centrality for
    the $\kt$ ranges 0.2--0.3, 0.3--0.4, 0.4--0.5, and 0.5--0.7~\gevc. The error
    bars indicate the statistical uncertainties and the square
    brackets show the systematic errors. The STAR data points, for 0--5\%, 5--10\%, 10--20\%, 20--30\% and 30--80\% Au--Au collisions, are slightly shifted for clarity.
  }
\label{fig:relative_Radii}
\end{figure}

Figure~\ref{fig:relative_Radii} shows the relative amplitudes of the
radius oscillations $R^{2}_{{\rm out},2}/R^{2}_{{\rm
    side},0}$,~$R^{2}_{{\rm side},2}/R^{2}_{{\rm side},0}$,
$R^{2}_{{\rm long},2}/R^{2}_{{\rm long},0}$, and $R^{2}_{{\rm
    os},2}/R^{2}_{{\rm side},0}$. When comparing our results to the
ones obtained by the STAR experiment, we observe similar relative
oscillations, however STAR results~\cite{Adams:2003ra,Adams:2004yc}
show on average larger oscillations for $\rside^{2}$. Our relative
amplitudes for $R^{2}_{\rm out,2}/R^{2}_{\rm side,0}$, $R^{2}_{\rm
  side,2}/R^{2}_{\rm side,0}$, and $R^{2}_{{\rm os},2}/R^{2}_{{\rm
    side},0}$ show a clear centrality dependence, whereas the $R^{2}_{\rm long,2}/R^{2}_{\rm long,0}$
is very close to zero for all centralities, similarly to the results
from RHIC~\cite{Adams:2003ra,Adamczyk:2014mxp,Adare:2015bcj}.

The source eccentricity is usually defined as
$\eps=(R^2_y-R^2_x)/(R^2_y+R^2_x)$, where $R_{x}$ is the in-plane
radius of the (assumed) elliptical source and $R_y$ is the
out-of-plane radius.  As shown in~\cite{Retiere:2003kf} the relative
amplitudes of side radii oscillations are mostly determined by the
spatial source anisotropy and are less affected by dynamical effects
such as velocity gradients.  The source eccentricity at freeze-out
$\eps_{\rm final}$ can be estimated from $R^{2}_{\rm side}$
oscillations at small pion momenta with an accuracy within 20--30\% as
$\eps_{\rm final}\approx
2\rsidetwo^2/\rsidez^2$~\cite{Retiere:2003kf}.

\begin{figure}[!ht]
\centering
\includegraphics[width=11cm,keepaspectratio]
{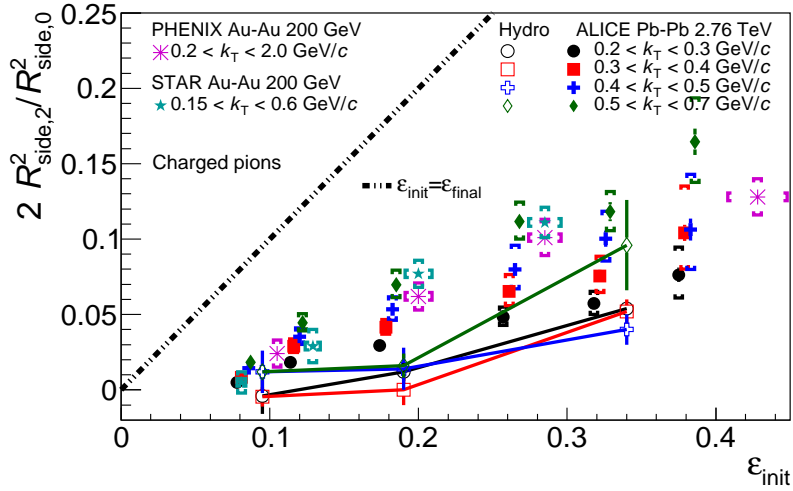}
  \caption{An estimate of freeze-out eccentricity $2
    R^{2}_{\rm side,2}/R^{2}_{\rm side,0}$ for different $\kt$ ranges vs.
    initial state eccentricity from Monte Carlo Glauber
    model~\cite{Ghosh:2016npo} for six centrality ranges, 0--5\%,
    5--10\%, 10--20\%, 20--30 \%, 30--40 \%, and 40--50\%.  The dashed line
    indicates $\eps_{\rm final}=\eps_{\rm init}$. Square brackets
    indicate systematic errors. }
\label{fig:finaleccentricity}
\end{figure}

Figure~\ref{fig:finaleccentricity} presents $2\rsidetwo^2/\rsidez^2$
for different $\kt$ ranges as a function of the initial-state
eccentricity for six different centralities and four $\kt$ bins.  For
the initial eccentricity we have used the nucleon participant
eccentricity from the Monte Carlo Glauber model for both,
Au--Au~collisions at \snn~=200~GeV~\cite{Adare:2014vax} and
\pb~collision at \snn~= 2.76~TeV~\cite{Ghosh:2016npo}. Our results for
all $\kt$ bins are significantly below the values of the initial
eccentricity indicating a more intense expansion in the in-plane
direction. Due to relatively large uncertainties of the RHIC results
for narrow $\kt$ bins, we compare our results only to the average STAR
data~\cite{Adams:2003ra} in $0.15<\kt <0.6$~\gevc~ and to PHENIX
results~\cite{Adare:2014vax} corresponding to $0.2<\kt <2.0$~\gevc~
($\mean{\kt}=0.53$~\gevc). We find a smaller final-state anisotropy in
the LHC regime compared to RHIC energies. This trend is qualitatively
consistent with expectations from hydrodynamic and transport
models~\cite{Shen:2012us,Lisa:2011na}. The final-state eccentricity
remains positive also at the LHC, evidence of an out-of-plane
elongated source at freeze-out. In Fig.~\ref{fig:finaleccentricity},
we also compare our results to the 3+1D hydrodynamic
calculations~\cite{Bozek:2014hwa}, which were performed for similar
centralities and $\kt$ ranges as in the experiment. This model
slightly underestimates the final source eccentricity.

%\clearpage

%=============================================================
%\section{Summary}

In conclusion, we have performed a measurement of two-pion azimuthally
differential femtoscopy relative to the second harmonic flow plane in
\pb~collisions at \snn~= 2.76~TeV.  The out, side, and out-side radii
exhibit clear oscillations while the long radius is consistent with a
constant.  The relative amplitudes of oscillations only weakly depend
on $\kt$, with the side-radii oscillation slightly increasing with
$\kt$.  The final-state source eccentricity, estimated via side-radius
oscillations, is noticeably smaller than at lower collisions energies,
but still exhibits an out-of-plane elongated source at freeze-out even
after a stronger in-plane expansion. The final eccentricity is
slightly larger than that predicted by existing hydrodynamic
calculations.

%%%%% acknowledgements
\newenvironment{acknowledgement}{\relax}{\relax}
\begin{acknowledgement}
\section*{Acknowledgements}
% Version: 2016-12-30

The ALICE Collaboration would like to thank all its engineers and technicians for their invaluable contributions to the construction of the experiment and the CERN accelerator teams for the outstanding performance of the LHC complex.
The ALICE Collaboration gratefully acknowledges the resources and support provided by all Grid centres and the Worldwide LHC Computing Grid (WLCG) collaboration.
The ALICE Collaboration acknowledges the following funding agencies for their support in building and running the ALICE detector:
A. I. Alikhanyan National Science Laboratory (Yerevan Physics Institute) Foundation (ANSL), State Committee of Science and World Federation of Scientists (WFS), Armenia;
Austrian Academy of Sciences and Nationalstiftung f\"{u}r Forschung, Technologie und Entwicklung, Austria;
Ministry of Communications and High Technologies, National Nuclear Research Center, Azerbaijan;
Conselho Nacional de Desenvolvimento Cient\'{\i}fico e Tecnol\'{o}gico (CNPq), Universidade Federal do Rio Grande do Sul (UFRGS), Financiadora de Estudos e Projetos (Finep) and Funda\c{c}\~{a}o de Amparo \`{a} Pesquisa do Estado de S\~{a}o Paulo (FAPESP), Brazil;
Ministry of Science \& Technology of China (MSTC), National Natural Science Foundation of China (NSFC) and Ministry of Education of China (MOEC) , China;
Ministry of Science, Education and Sport and Croatian Science Foundation, Croatia;
Ministry of Education, Youth and Sports of the Czech Republic, Czech Republic;
The Danish Council for Independent Research | Natural Sciences, the Carlsberg Foundation and Danish National Research Foundation (DNRF), Denmark;
Helsinki Institute of Physics (HIP), Finland;
Commissariat \`{a} l'Energie Atomique (CEA) and Institut National de Physique Nucl\'{e}aire et de Physique des Particules (IN2P3) and Centre National de la Recherche Scientifique (CNRS), France;
Bundesministerium f\"{u}r Bildung, Wissenschaft, Forschung und Technologie (BMBF) and GSI Helmholtzzentrum f\"{u}r Schwerionenforschung GmbH, Germany;
Ministry of Education, Research and Religious Affairs, Greece;
National Research, Development and Innovation Office, Hungary;
Department of Atomic Energy Government of India (DAE) and Council of Scientific and Industrial Research (CSIR), New Delhi, India;
Indonesian Institute of Science, Indonesia;
Centro Fermi - Museo Storico della Fisica e Centro Studi e Ricerche Enrico Fermi and Istituto Nazionale di Fisica Nucleare (INFN), Italy;
Institute for Innovative Science and Technology , Nagasaki Institute of Applied Science (IIST), Japan Society for the Promotion of Science (JSPS) KAKENHI and Japanese Ministry of Education, Culture, Sports, Science and Technology (MEXT), Japan;
Consejo Nacional de Ciencia (CONACYT) y Tecnolog\'{i}a, through Fondo de Cooperaci\'{o}n Internacional en Ciencia y Tecnolog\'{i}a (FONCICYT) and Direcci\'{o}n General de Asuntos del Personal Academico (DGAPA), Mexico;
Nationaal instituut voor subatomaire fysica (Nikhef), Netherlands;
The Research Council of Norway, Norway;
Commission on Science and Technology for Sustainable Development in the South (COMSATS), Pakistan;
Pontificia Universidad Cat\'{o}lica del Per\'{u}, Peru;
Ministry of Science and Higher Education and National Science Centre, Poland;
Korea Institute of Science and Technology Information and National Research Foundation of Korea (NRF), Republic of Korea;
Ministry of Education and Scientific Research, Institute of Atomic Physics and Romanian National Agency for Science, Technology and Innovation, Romania;
Joint Institute for Nuclear Research (JINR), Ministry of Education and Science of the Russian Federation and National Research Centre Kurchatov Institute, Russia;
Ministry of Education, Science, Research and Sport of the Slovak Republic, Slovakia;
National Research Foundation of South Africa, South Africa;
Centro de Aplicaciones Tecnol\'{o}gicas y Desarrollo Nuclear (CEADEN), Cubaenerg\'{\i}a, Cuba, Ministerio de Ciencia e Innovacion and Centro de Investigaciones Energ\'{e}ticas, Medioambientales y Tecnol\'{o}gicas (CIEMAT), Spain;
Swedish Research Council (VR) and Knut \& Alice Wallenberg Foundation (KAW), Sweden;
European Organization for Nuclear Research, Switzerland;
National Science and Technology Development Agency (NSDTA), Suranaree University of Technology (SUT) and Office of the Higher Education Commission under NRU project of Thailand, Thailand;
Turkish Atomic Energy Agency (TAEK), Turkey;
National Academy of  Sciences of Ukraine, Ukraine;
Science and Technology Facilities Council (STFC), United Kingdom;
National Science Foundation of the United States of America (NSF) and United States Department of Energy, Office of Nuclear Physics (DOE NP), United States of America.
\end{acknowledgement}

\bibliography{reference}
\bibliographystyle{utphys}

%%%%%%%%% appendix with author list
\newpage
\appendix

\section{The ALICE Collaboration}
\label{app:collab}

% Collaboration: CERN-LHC-ALICE
% Generation Date is 2016-Dec-30

% How to use:
%%%%%%%%% appendix with author list
%\appendix
%\section{The ALICE Collaboration}
%\label{app:collab}
%\input{authors-list.tex}  %%%%%%% get the latest version before submitting

\begingroup
\small
\begin{flushleft}
D.~Adamov\'{a}$^\textrm{\scriptsize 87}$,
M.M.~Aggarwal$^\textrm{\scriptsize 91}$,
G.~Aglieri Rinella$^\textrm{\scriptsize 34}$,
M.~Agnello$^\textrm{\scriptsize 30}$\textsuperscript{,}$^\textrm{\scriptsize 113}$,
N.~Agrawal$^\textrm{\scriptsize 47}$,
Z.~Ahammed$^\textrm{\scriptsize 139}$,
S.~Ahmad$^\textrm{\scriptsize 17}$,
S.U.~Ahn$^\textrm{\scriptsize 69}$,
S.~Aiola$^\textrm{\scriptsize 143}$,
A.~Akindinov$^\textrm{\scriptsize 54}$
% ...
\renewcommand\labelenumi{\textsuperscript{\theenumi}~}

\section*{Affiliation notes}
\renewcommand\theenumi{\roman{enumi}}
\begin{Authlist}
\item \Adef{0}Deceased
\item \Adef{idp1764384}{Also at: Georgia State University, Atlanta, Georgia, United States}
\item \Adef{idp3207120}{Also at: Also at Department of Applied Physics, Aligarh Muslim University, Aligarh, India}
\item \Adef{idp3984928}{Also at: M.V. Lomonosov Moscow State University, D.V. Skobeltsyn Institute of Nuclear, Physics, Moscow, Russia}
\end{Authlist}

\section*{Collaboration Institutes}
\renewcommand\theenumi{\arabic{enumi}~}

$^{1}$A.I. Alikhanyan National Science Laboratory (Yerevan Physics Institute) Foundation, Yerevan, Armenia
\\
$^{2}$Benem\'{e}rita Universidad Aut\'{o}noma de Puebla, Puebla, Mexico
\\
$^{3}$Bogolyubov Institute for Theoretical Physics, Kiev, Ukraine
\\
$^{4}$Bose Institute, Department of Physics 
and Centre for Astroparticle Physics and Space Science (CAPSS), Kolkata, India
\\
$^{5}$Budker Institute for Nuclear Physics, Novosibirsk, Russia
\\
$^{6}$California Polytechnic State University, San Luis Obispo, California, United States
\\
$^{7}$Central China Normal University, Wuhan, China
\\
$^{8}$Centre de Calcul de l'IN2P3, Villeurbanne, Lyon, France
\\
$^{9}$Centro de Aplicaciones Tecnol\'{o}gicas y Desarrollo Nuclear (CEADEN), Havana, Cuba
\\
$^{10}$Centro de Investigaciones Energ\'{e}ticas Medioambientales y Tecnol\'{o}gicas (CIEMAT), Madrid, Spain
\\
$^{11}$Centro de Investigaci\'{o}n y de Estudios Avanzados (CINVESTAV), Mexico City and M\'{e}rida, Mexico
\\
$^{12}$Centro Fermi - Museo Storico della Fisica e Centro Studi e Ricerche ``Enrico Fermi', Rome, Italy
\\
$^{13}$Chicago State University, Chicago, Illinois, United States
\\
$^{14}$China Institute of Atomic Energy, Beijing, China
\\
$^{15}$COMSATS Institute of Information Technology (CIIT), Islamabad, Pakistan
\\
$^{16}$Departamento de F\'{\i}sica de Part\'{\i}culas and IGFAE, Universidad de Santiago de Compostela, Santiago de Compostela, Spain
\\
$^{17}$Department of Physics, Aligarh Muslim University, Aligarh, India
\\
$^{18}$Department of Physics, Ohio State University, Columbus, Ohio, United States
\\
$^{19}$Department of Physics, Sejong University, Seoul, South Korea
\\
$^{20}$Department of Physics, University of Oslo, Oslo, Norway
\\
$^{21}$Department of Physics and Technology, University of Bergen, Bergen, Norway
\\
$^{22}$Dipartimento di Fisica dell'Universit\`{a} 'La Sapienza'
and Sezione INFN, Rome, Italy
\\
$^{23}$Dipartimento di Fisica dell'Universit\`{a}
and Sezione INFN, Cagliari, Italy
\\
$^{24}$Dipartimento di Fisica dell'Universit\`{a}
and Sezione INFN, Trieste, Italy
\\
$^{25}$Dipartimento di Fisica dell'Universit\`{a}
and Sezione INFN, Turin, Italy
\\
$^{26}$Dipartimento di Fisica e Astronomia dell'Universit\`{a}
and Sezione INFN, Bologna, Italy
\\
$^{27}$Dipartimento di Fisica e Astronomia dell'Universit\`{a}
and Sezione INFN, Catania, Italy
\\
$^{28}$Dipartimento di Fisica e Astronomia dell'Universit\`{a}
and Sezione INFN, Padova, Italy
\\
$^{29}$Dipartimento di Fisica `E.R.~Caianiello' dell'Universit\`{a}
and Gruppo Collegato INFN, Salerno, Italy
\\
$^{30}$Dipartimento DISAT del Politecnico and Sezione INFN, Turin, Italy
\\
$^{31}$Dipartimento di Scienze e Innovazione Tecnologica dell'Universit\`{a} del Piemonte Orientale and INFN Sezione di Torino, Alessandria, Italy
\\
$^{32}$Dipartimento Interateneo di Fisica `M.~Merlin'
and Sezione INFN, Bari, Italy
\\
$^{33}$Division of Experimental High Energy Physics, University of Lund, Lund, Sweden
\\
$^{34}$European Organization for Nuclear Research (CERN), Geneva, Switzerland
\\
$^{35}$Excellence Cluster Universe, Technische Universit\"{a}t M\"{u}nchen, Munich, Germany
\\
$^{36}$Faculty of Engineering, Bergen University College, Bergen, Norway
\\
$^{37}$Faculty of Mathematics, Physics and Informatics, Comenius University, Bratislava, Slovakia
\\
$^{38}$Faculty of Nuclear Sciences and Physical Engineering, Czech Technical University in Prague, Prague, Czech Republic
\\
$^{39}$Faculty of Science, P.J.~\v{S}af\'{a}rik University, Ko\v{s}ice, Slovakia
\\
$^{40}$Faculty of Technology, Buskerud and Vestfold University College, Tonsberg, Norway
\\
$^{41}$Frankfurt Institute for Advanced Studies, Johann Wolfgang Goethe-Universit\"{a}t Frankfurt, Frankfurt, Germany
\\
$^{42}$Gangneung-Wonju National University, Gangneung, South Korea
\\
$^{43}$Gauhati University, Department of Physics, Guwahati, India
\\
$^{44}$Helmholtz-Institut f\"{u}r Strahlen- und Kernphysik, Rheinische Friedrich-Wilhelms-Universit\"{a}t Bonn, Bonn, Germany
\\
$^{45}$Helsinki Institute of Physics (HIP), Helsinki, Finland
\\
$^{46}$Hiroshima University, Hiroshima, Japan
\\
$^{47}$Indian Institute of Technology Bombay (IIT), Mumbai, India
\\
$^{48}$Indian Institute of Technology Indore, Indore, India
\\
$^{49}$Indonesian Institute of Sciences, Jakarta, Indonesia
\\
$^{50}$Inha University, Incheon, South Korea
\\
$^{51}$Institut de Physique Nucl\'eaire d'Orsay (IPNO), Universit\'e Paris-Sud, CNRS-IN2P3, Orsay, France
\\
$^{52}$Institute for Nuclear Research, Academy of Sciences, Moscow, Russia
\\
$^{53}$Institute for Subatomic Physics of Utrecht University, Utrecht, Netherlands
\\
$^{54}$Institute for Theoretical and Experimental Physics, Moscow, Russia
\\
$^{55}$Institute of Experimental Physics, Slovak Academy of Sciences, Ko\v{s}ice, Slovakia
\\
$^{56}$Institute of Physics, Academy of Sciences of the Czech Republic, Prague, Czech Republic
\\
$^{57}$Institute of Physics, Bhubaneswar, India
\\
$^{58}$Institute of Space Science (ISS), Bucharest, Romania
\\
$^{59}$Institut f\"{u}r Informatik, Johann Wolfgang Goethe-Universit\"{a}t Frankfurt, Frankfurt, Germany
\\
$^{60}$Institut f\"{u}r Kernphysik, Johann Wolfgang Goethe-Universit\"{a}t Frankfurt, Frankfurt, Germany
\\
$^{61}$Institut f\"{u}r Kernphysik, Westf\"{a}lische Wilhelms-Universit\"{a}t M\"{u}nster, M\"{u}nster, Germany
\\
$^{62}$Instituto de Ciencias Nucleares, Universidad Nacional Aut\'{o}noma de M\'{e}xico, Mexico City, Mexico
\\
$^{63}$Instituto de F\'{i}sica, Universidade Federal do Rio Grande do Sul (UFRGS), Porto Alegre, Brazil
\\
$^{64}$Instituto de F\'{\i}sica, Universidad Nacional Aut\'{o}noma de M\'{e}xico, Mexico City, Mexico
\\
$^{65}$IRFU, CEA, Universit\'{e} Paris-Saclay, F-91191 Gif-sur-Yvette, France, Saclay, France
\\
$^{66}$iThemba LABS, National Research Foundation, Somerset West, South Africa
\\
$^{67}$Joint Institute for Nuclear Research (JINR), Dubna, Russia
\\
$^{68}$Konkuk University, Seoul, South Korea
\\
$^{69}$Korea Institute of Science and Technology Information, Daejeon, South Korea
\\
$^{70}$KTO Karatay University, Konya, Turkey
\\
$^{71}$Laboratoire de Physique Corpusculaire (LPC), Clermont Universit\'{e}, Universit\'{e} Blaise Pascal, CNRS--IN2P3, Clermont-Ferrand, France
\\
$^{72}$Laboratoire de Physique Subatomique et de Cosmologie, Universit\'{e} Grenoble-Alpes, CNRS-IN2P3, Grenoble, France
\\
$^{73}$Laboratori Nazionali di Frascati, INFN, Frascati, Italy
\\
$^{74}$Laboratori Nazionali di Legnaro, INFN, Legnaro, Italy
\\
$^{75}$Lawrence Berkeley National Laboratory, Berkeley, California, United States
\\
$^{76}$Moscow Engineering Physics Institute, Moscow, Russia
\\
$^{77}$Nagasaki Institute of Applied Science, Nagasaki, Japan
\\
$^{78}$National and Kapodistrian University of Athens, Physics Department, Athens, Greece, Athens, Greece
\\
$^{79}$National Centre for Nuclear Studies, Warsaw, Poland
\\
$^{80}$National Institute for Physics and Nuclear Engineering, Bucharest, Romania
\\
$^{81}$National Institute of Science Education and Research, Bhubaneswar, India
\\
$^{82}$National Nuclear Research Center, Baku, Azerbaijan
\\
$^{83}$National Research Centre Kurchatov Institute, Moscow, Russia
\\
$^{84}$Niels Bohr Institute, University of Copenhagen, Copenhagen, Denmark
\\
$^{85}$Nikhef, Nationaal instituut voor subatomaire fysica, Amsterdam, Netherlands
\\
$^{86}$Nuclear Physics Group, STFC Daresbury Laboratory, Daresbury, United Kingdom
\\
$^{87}$Nuclear Physics Institute, Academy of Sciences of the Czech Republic, \v{R}e\v{z} u Prahy, Czech Republic
\\
$^{88}$Oak Ridge National Laboratory, Oak Ridge, Tennessee, United States
\\
$^{89}$Petersburg Nuclear Physics Institute, Gatchina, Russia
\\
$^{90}$Physics Department, Creighton University, Omaha, Nebraska, United States
\\
$^{91}$Physics Department, Panjab University, Chandigarh, India
\\
$^{92}$Physics Department, University of Cape Town, Cape Town, South Africa
\\
$^{93}$Physics Department, University of Jammu, Jammu, India
\\
$^{94}$Physics Department, University of Rajasthan, Jaipur, India
\\
$^{95}$Physikalisches Institut, Eberhard Karls Universit\"{a}t T\"{u}bingen, T\"{u}bingen, Germany
\\
$^{96}$Physikalisches Institut, Ruprecht-Karls-Universit\"{a}t Heidelberg, Heidelberg, Germany
\\
$^{97}$Physik Department, Technische Universit\"{a}t M\"{u}nchen, Munich, Germany
\\
$^{98}$Purdue University, West Lafayette, Indiana, United States
\\
$^{99}$Pusan National University, Pusan, South Korea
\\
$^{100}$Research Division and ExtreMe Matter Institute EMMI, GSI Helmholtzzentrum f\"ur Schwerionenforschung GmbH, Darmstadt, Germany
\\
$^{101}$Rudjer Bo\v{s}kovi\'{c} Institute, Zagreb, Croatia
\\
$^{102}$Russian Federal Nuclear Center (VNIIEF), Sarov, Russia
\\
$^{103}$Saha Institute of Nuclear Physics, Kolkata, India
\\
$^{104}$School of Physics and Astronomy, University of Birmingham, Birmingham, United Kingdom
\\
$^{105}$Secci\'{o}n F\'{\i}sica, Departamento de Ciencias, Pontificia Universidad Cat\'{o}lica del Per\'{u}, Lima, Peru
\\
$^{106}$Sezione INFN, Bari, Italy
\\
$^{107}$Sezione INFN, Bologna, Italy
\\
$^{108}$Sezione INFN, Cagliari, Italy
\\
$^{109}$Sezione INFN, Catania, Italy
\\
$^{110}$Sezione INFN, Padova, Italy
\\
$^{111}$Sezione INFN, Rome, Italy
\\
$^{112}$Sezione INFN, Trieste, Italy
\\
$^{113}$Sezione INFN, Turin, Italy
\\
$^{114}$SSC IHEP of NRC Kurchatov institute, Protvino, Russia
\\
$^{115}$Stefan Meyer Institut f\"{u}r Subatomare Physik (SMI), Vienna, Austria
\\
$^{116}$SUBATECH, Ecole des Mines de Nantes, Universit\'{e} de Nantes, CNRS-IN2P3, Nantes, France
\\
$^{117}$Suranaree University of Technology, Nakhon Ratchasima, Thailand
\\
$^{118}$Technical University of Ko\v{s}ice, Ko\v{s}ice, Slovakia
\\
$^{119}$Technical University of Split FESB, Split, Croatia
\\
$^{120}$The Henryk Niewodniczanski Institute of Nuclear Physics, Polish Academy of Sciences, Cracow, Poland
\\
$^{121}$The University of Texas at Austin, Physics Department, Austin, Texas, United States
\\
$^{122}$Universidad Aut\'{o}noma de Sinaloa, Culiac\'{a}n, Mexico
\\
$^{123}$Universidade de S\~{a}o Paulo (USP), S\~{a}o Paulo, Brazil
\\
$^{124}$Universidade Estadual de Campinas (UNICAMP), Campinas, Brazil
\\
$^{125}$Universidade Federal do ABC, Santo Andre, Brazil
\\
$^{126}$University of Houston, Houston, Texas, United States
\\
$^{127}$University of Jyv\"{a}skyl\"{a}, Jyv\"{a}skyl\"{a}, Finland
\\
$^{128}$University of Liverpool, Liverpool, United Kingdom
\\
$^{129}$University of Tennessee, Knoxville, Tennessee, United States
\\
$^{130}$University of the Witwatersrand, Johannesburg, South Africa
\\
$^{131}$University of Tokyo, Tokyo, Japan
\\
$^{132}$University of Tsukuba, Tsukuba, Japan
\\
$^{133}$University of Zagreb, Zagreb, Croatia
\\
$^{134}$Universit\'{e} de Lyon, Universit\'{e} Lyon 1, CNRS/IN2P3, IPN-Lyon, Villeurbanne, Lyon, France
\\
$^{135}$Universit\'{e} de Strasbourg, CNRS, IPHC UMR 7178, F-67000 Strasbourg, France, Strasbourg, France
\\
$^{136}$Universit\`{a} degli Studi di Pavia, Pavia, Italy
\\
$^{137}$Universit\`{a} di Brescia, Brescia, Italy
\\
$^{138}$V.~Fock Institute for Physics, St. Petersburg State University, St. Petersburg, Russia
\\
$^{139}$Variable Energy Cyclotron Centre, Kolkata, India
\\
$^{140}$Warsaw University of Technology, Warsaw, Poland
\\
$^{141}$Wayne State University, Detroit, Michigan, United States
\\
$^{142}$Wigner Research Centre for Physics, Hungarian Academy of Sciences, Budapest, Hungary
\\
$^{143}$Yale University, New Haven, Connecticut, United States
\\
$^{144}$Yonsei University, Seoul, South Korea
\\
$^{145}$Zentrum f\"{u}r Technologietransfer und Telekommunikation (ZTT), Fachhochschule Worms, Worms, Germany
\endgroup

  %%%%%%% done by webmaster team
\end{document}